\documentclass[a4paper]{jpconf}
\usepackage{graphicx}
\usepackage{citesort}
\begin{document}
\title{The ASY-EOS experiment at GSI: investigating the symmetry energy at supra-saturation densities}
%
%
\author{P.~Russotto$^{1}$, M.~Chartier$^{2}$, E.~De~Filippo$^{1}$, A.~Le~F\`{e}vre$^{3}$, S.~Gannon$^{2}$,
I.~Ga\v{s}pari\'c$^{4,5}$, M.~Ki\v{s}$^{3,4}$, S.~Kupny$^{6}$, Y.~Leifels$^{3}$, R.C.~Lemmon$^{7}$, J.~{\L}ukasik$^{8}$, P.~Marini$^{9,10}$, A.~Pagano$^{1}$, P.~Paw{\l}owski$^{8}$, S.~Santoro$^{11,12}$, W.~Trautmann$^{3}$, M.~Veselsky$^{13}$, L.~Acosta$^{14}$, M.~Adamczyk$^{6}$, A.~Al-Ajlan$^{15}$, M.~Al-Garawi$^{16}$, S.~Al-Homaidhi$^{15}$, F.~Amorini$^{14}$, L.~Auditore$^{11,12}$, T.~Aumann$^{5}$, Y.~Ayyad$^{17}$, V.~Baran$^{14,18}$, Z.~Basrak$^{4}$, J.~Benlliure$^{17}$, C.~Boiano$^{19}$, M.~Boisjoli$^{10}$, K.~Boretzky$^{3}$, J.~Brzychczyk$^{6}$, A.~Budzanowski$^{8}$, G.~Cardella$^{1}$, P.~Cammarata$^{9}$, Z.~Chajecki$^{20}$, A.~Chbihi$^{10}$, M.~Colonna$^{14}$, D.~Cozma$^{21}$, B.~Czech$^{8}$, M.~Di~Toro$^{14,22}$, M.~Famiano$^{23}$, E.~Geraci$^{1,22}$, V.~Greco$^{14,22}$, L.~Grassi$^{4}$, C.~Guazzoni$^{19,24}$, P.~Guazzoni$^{19,25}$, M.~Heil$^{3}$, L.~Heilborn$^{9}$, R.~Introzzi$^{26}$, T.~Isobe$^{27}$, K.~Kezzar$^{16}$, A.~Krasznahorkay$^{28}$, N.~Kurz$^{3}$, E.~La~Guidara$^{1}$, G.~Lanzalone$^{14,29}$, P.~Lasko$^{6}$, Q.~Li$^{30}$, I.~Lombardo$^{30,31}$, W.~G.~Lynch$^{20}$, Z.~Matthews$^{2}$, L.~May$^{9}$, T.~Minniti$^{11,12}$, M.~Mostazo$^{17}$, M.~Papa$^{1}$, S.~Pirrone$^{1}$, G.~Politi$^{1,22}$, F.~Porto$^{14,22}$, R.~Reifarth$^{3}$, W.~Reisdorf$^{3}$, F.~Riccio$^{19,25}$, F.~Rizzo$^{14,22}$, E.~Rosato$^{30,31}$, D.~Rossi$^{3}$, H.~Simon$^{3}$, I.~Skwirczynska$^{8}$, Z.~Sosin$^{6}$, L.~Stuhl$^{28}$, A.~Trifir\`{o}$^{11,12}$,M.~Trimarchi$^{11,12}$,M.~B.~Tsang$^{20}$,  G.~Verde$^{1}$, M.~Vigilante$^{30,31}$, A.~Wieloch$^{6}$, P.~Wigg$^{2}$, H.~H.~Wolter$^{32}$, P.~Wu$^{2}$, S.~Yennello$^{9}$,
P.~Zambon$^{19,24}$, L.~Zetta$^{19,25}$ and M.~Zoric$^{4}$
}
\address{$^{1}$INFN-Sezione di Catania, Catania, Italy}
\address{$^{2}$University of Liverpool, Liverpool, UK}
\address{$^{3}$GSI Helmholtzzentrum, Darmstadt, Germany}
\address{$^{4}$Ruder Bo\v{s}kovi\`{c} Institute, Zagreb, Croatia}
\address{$^{5}$Technische Universit\"{a}t, Darmstadt, Germany }
\address{$^{6}$Jagiellonian University, Krak\`{o}w, Poland}
\address{$^{7}$STFC Laboratory, Daresbury, UK}
\address{$^{8}$IFJ-PAN, Krakow, Poland }
\address{$^{9}$Texas A$\&$M University, College Station, USA }
\address{$^{10}$GANIL, Caen, France }
\address{$^{11}$INFN-Gruppo Collegato di Messina, Messina, Italy}
\address{$^{12}$Universit\`{a} di Messina, Messina, Italy }
\address{$^{13}$Institute of Physics, Slovak Academy of Sciences, Bratislava, Slovakia}
\address{$^{14}$INFN-Laboratori Nazionali del Sud, Catania, Italy}
\address{$^{15}$KACST Riyadh, Riyadh, Saudi Arabia}
\address{$^{16}$King Saud University, Riyadh, Saudi Arabia }
\address{$^{17}$University of Santiago de Compostela, Santiago de Compostela, Spain}
\address{$^{18}$University of Bucharest, Bucharest, Romania}
\address{$^{19}$INFN-Sezione di Milano, Milano, Ialy }
\address{$^{20}$NSCL Michigan State University, East Lansing, USA }
\address{$^{21}$IFIN-HH, Magurele-Bucharest,Romania}
\address{$^{22}$Universit\`{a} di Catania, Catania, Italy}
\address{$^{21}$Western Michigan University, USA }
\address{$^{24}$Politecnico di Milano, Milano, Italy }
\address{$^{25}$Universit\`{a} degli Studi di Milano, Milano, Italy }
\address{$^{26}$INFN, Politecnico di Torino, Torino, Italy }
\address{$^{27}$RIKEN, Wako, Japan}
\address{$^{28}$Institute of Nuclear Research, Debrecen, Hungary}
\address{$^{29}$Universit\`{a} Kore, Enna, Italy}
\address{$^{30}$Huzhou Teachers College, China }
\address{$^{30}$INFN-Sezione di Napoli, Napoli, Italy }
\address{$^{31}$Universit\`{a} di Napoli, Napoli, Italy }
\address{$^{32}$LMU, M\"{u}nchen, Germany}
\ead{russotto@lns.infn.it}
\begin{abstract}
The elliptic-flow ratio of neutrons with respect to protons in reactions of neutron rich heavy-ions systems at
intermediate energies has been proposed as an observable sensitive to the strength of the symmetry term in the
nuclear Equation Of State (EOS) at supra-saturation densities. The recent results obtained from the existing
FOPI/LAND data for $^{197}$Au+$^{197}$Au collisions at 400 MeV/nucleon in comparison with the UrQMD model
allowed a first estimate of the symmetry term of the EOS but suffer from a considerable statistical
uncertainty.
In order to obtain an improved data set for Au+Au collisions and to extend the study to other systems, a new
experiment was carried out at the GSI laboratory by the ASY-EOS collaboration in May 2011.
\end{abstract}
\section{Introduction}
A key question in modern nuclear physics is the knowledge of the nuclear Equation Of State (EOS) and, in
particular, of its dependence on density and on asymmetry, i.e., on the relative neutron-to-proton abundance. The
EOS can be divided into a symmetric term (i.e., independent of the isospin asymmetry I=$\frac{N-Z}{N+Z}$, where
N and Z are the numbers of neutrons and protons, respectively) and an asymmetric term given by the symmetry energy, i.e. the difference between the EOS for neutron matter and for symmetric one, multiplied by the square of the isospin asymmetry I~\cite{Dan02,Lat04,Bar05,Fuc06,Bao08}.\\
Measurements of isoscalar collective vibrations, collective flow and kaon production~\cite{You99,Dan02,Fuc06} in
energetic nucleus-nucleus collisions have constrained the behavior of the EOS of isospin symmetric matter for
densities up to five times the saturation density $\rho_{0}$. On the other side, the EOS of asymmetric matter is
still subject to large uncertainties. Besides the astrophysical interest, e.g. neutron star physics and supernovae
dynamics~\cite{Lat00,Bot04}, the density dependence of the symmetry energy is of fundamental importance for nuclear
physics. The thickness of the neutron skin of heavy nuclei reflects the differential pressure exerted on the
core~\cite{Hor01} and the strength of the three-body forces, an important ingredient in nuclear structure
calculations~\cite{Wir02}, represents one of the major uncertainties in modeling the EOS at high
density~\cite{Fuc06,Cha10}. Moreover, properties of nuclei away from the valley of stability  and the dynamics of
nuclear reactions at Fermi energies rely on the density dependence of the symmetry energy~\cite{Bar05,Bao08}.\\
In the last decade, measurements of the Giant Monopole~\cite{Li07}, Giant Dipole~\cite{Tri08} and Pygmy
Dipole~\cite{Kli07} resonances in neutron-rich nuclei, isospin diffusion~\cite{Tsa04,Tsa09}, neutron and proton
emissions~\cite{Fam06}, fragment isotopic ratios~\cite{Tsa09,Igl06,Tsa01}, isospin dependence of competition
between deep-inelastic and incomplete fusion reactions~\cite{Amo09}, neutron enrichment of mid-velocity
fragments~\cite{Def12} have provided constraints on the density dependence of the symmetry energy around and below
saturation density $\rho_{0}$~\cite{Tsa12}. It results that the best description of experimental data is obtained
with a symmetry energy $S(u)=C^{sym}_{kin}(u)^{2/3}+C^{sym}_{pot}(u)^{\gamma}$ with ${\gamma}$ in the range
0.6-1.1 ($u=\rho/\rho_{0}$
is the reduced nuclear density). In the near future, extensions of these measurements with both stable and
rare-isotope beams will provide further stringent constraints at sub-saturation densities.\\
In contrast, up to now, very few experimental constraints exist on the symmetry energy at supra-saturation
densities ($u>1$). This is the domain with the greatest theoretical uncertainty and the largest interest
for neutron stars. The behavior of the asymmetric EOS at supra-saturation densities can only be explored in
terrestrial
laboratories by using relativistic heavy-ion collisions of isospin asymmetric nuclei. Reaction simulations propose
several potentially useful observable such as neutron and proton flows (direct and
elliptic)~\cite{Gre03,Yon06ab,Bar05,Bao08}, neutron/proton ratio~\cite{Li06a,Li06b,Bao08,Tsa09,Kum12,Fen12},
$\pi^{-}/\pi^{+}$ ratio and flows~\cite{Li05a,Bar05,Yon06ab,Bao08,Fen12}, $K^{+}/K^{0}$ ~\cite{Fer06} and
$\Sigma^{-}$/$\Sigma^{+}$~\cite{Li05a} ratios, flow and yield of light charged particles, with special emphasis on
isotope pairs (e.g. $^{3}H$ and $^{3}He$, $^{7}Li$ and $^{7}Be$)~\cite{Bar05,Bao08,Kum12,Fen12}. To this day the
problem is still open, since few studies have provided constraints on the behavior of the symmetry energy at supra-saturation
densities (see \cite{Tra12} for a review).\\
The single ratio $\pi^{-}/\pi^{+}$ was measured in $^{197}Au+^{197}Au$~\cite{Rei07} and analyzed using the
hadronic transport model IBUU04~\cite{Xia09}. The results suggest that the symmetry energy is rather soft at
supra-saturation densities; this finding - symmetry energy reaches its maximum at a density between $\rho_{0}$ and
$2\rho_{0}$ and then starts decreasing at higher densities - is not consistent with most of the extrapolation near
or below saturation density.
The same set of FOPI data has been analyzed in the framework of the IMproved Isospin dependent Quantum Molecular
Dynamics (ImIQMD)~\cite{Fen10}; it results in a very stiff symmetry energy of the potential term proportional to
$u^{\gamma}$ with ${\gamma}\simeq 2$. Moreover, work~\cite{Li06c} suggests a reduced sensitivity
of the $\pi^{-}/\pi^{+}$ ratio to the symmetry energy. It follows that for the pion signal further work is
needed to establish the effectiveness in probing the symmetry energy. In-medium absorption and re-emission of
pions can distort the asymptotic experimental signal and it is not clear which density of matter is explored by
the pions signal. The analysis of another set of FOPI data is described in the next section of this paper.\\
In addition, we want to remind the reader that the FOPI Collaboration has recently published a complete systematic of direct and
elliptic flows for Light Charged Particles (LCP) for 25 combinations of systems and energies~\cite{Rei12}, that
represents an unique set of data to study symmetry energy effects in flow and yield of LCP.
\section{Neutron and proton elliptic flows} One of the most promising probe of the symmetry energy strength at
supra-saturation densities is the difference of the neutron and proton (or hydrogen) elliptic
flows~\cite{Tra09ab10,Rus11}. This has emerged firstly from calculations based on the Ultra-Relativistic Quantum
Molecular Dynamics model (UrQMD)~\cite{Urq}. We report here some results obtained using UrQMD for the
$^{197}Au+^{197}Au$ collision at 400 MeV/nucleon. The calculations have been performed using both Asy-Stiff
$(\gamma=1.5)$ and Asy-Soft $(\gamma=0.5)$ potential symmetry energies. The UrQMD predictions for the elliptic
flow of neutrons, protons and hydrogen as a function of rapidity in the laboratory frame, $Y_{lab}$, for
mid-peripheral collisions (impact parameter $5.5<b<7.5$ fm) and for the two choices of the density dependence of
the symmetry energy, are shown in Fig. 1.
\begin{figure}
\centering
\includegraphics[scale=.40]{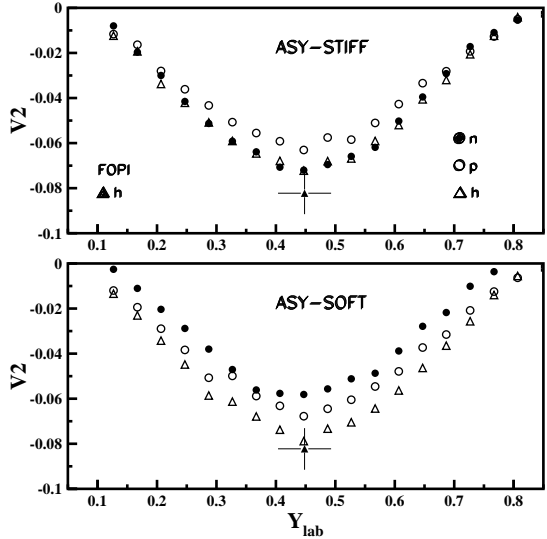}
\caption{Elliptic flow parameter $v_{2}$ for mid-peripheral (impact parameter $5.5<b<7.5$ fm) $^{197}Au+^{197}Au$
collisions at 400 MeV/nucleon as calculated with the UrQMD model for neutrons (dots), protons (circles), and all
hydrogen isotopes (Z=1, open triangles), integrated over transverse momentum $p_{t}$, as a function of the laboratory
rapidity $Y_{lab}$. The predictions obtained with a stiff and a soft density dependence of the symmetry term are
given in the upper and lower panels, respectively. The experimental result from Ref.~\cite{And05} for Z = 1
particles at mid-rapidity is represented by the filled triangle (the horizontal bar represents the experimental
rapidity interval); from ~\cite{Rus11}.}
\end{figure}
 We remind here that direct ($v_{1}$) and elliptic ($v_{2}$) flows are obtained by the azimuthal particle
 distributions with the usual Fourier expansion
\begin{equation}
\label{all}
 f(\Delta\phi) \propto 1 + 2 \cdot v_{1} \cdot cos(\Delta\phi) + 2 \cdot v_{2} \cdot cos(2 \cdot \Delta\phi)
\end{equation}
with $\Delta\phi$  representing the azimuthal angle of the emitted particle with respect to the reaction
plane~\cite{And08}. The dominant difference is the significantly larger neutron squeeze-out in the Asy-Stiff case
(upper panel) compared to the Asy-Soft case (lower panel).
Neutron and proton directed and elliptic flows were measured in $^{197}Au+^{197}Au$ collisions at 400 MeV/nucleon
using the
LAND neutron detector and the FOPI Phase 1 forward wall~\cite{Lei93}. We have reanalyzed the data to extract the
neutron and proton direct and elliptic flows~\cite{Tra09ab10,Rus11}. The results have been compared with
predictions of the UrQMD model
 with the aim of providing constraints on the symmetry energy at supra-saturation densities. The dependence of the
 elliptic flow parameter $v_{2}$ on the transverse momentum per nucleon, $p_{t}/A$, is shown in Fig. 2, upper panel,
 for the combined data set of collisions with impact parameter $b<7.5$ fm; the impact parameter has been estimated
 from the total charged particle multiplicity registered in the FOPI forward wall.
\begin{figure}
\centering
\includegraphics[scale=.33]{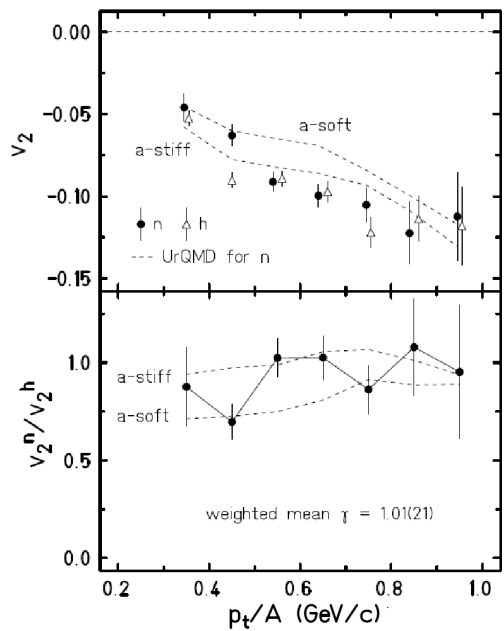}
\caption{Differential elliptic flow parameters $v_{2}$ for neutrons (dots) and hydrogen isotopes (open triangles,
top panel) and their
ratio (lower panels) for moderately central ($b < 7.5 fm$) collisions of $^{197}Au+^{197}Au$ at 400 MeV/nucleon,
as a function
of the transverse momentum per nucleon $p_{t}/A$. The symbols represent the experimental data. The UrQMD predictions
for $\gamma$ = 1.5
(Asy-Stiff) and $\gamma$= 0.5 (Asy-Soft) obtained for neutrons (top panel) and for the ratio (bottom panel) are
given by the dashed
lines; adapted from ~\cite{Rus11}.}
\end{figure}
For the quantitative evaluation, the ratio of the flow parameters of neutrons versus hydrogen isotopes has been
used. The results for the ratio $v^{n}_{2}/v^{h}_{2}$, i.e. with respect to the integrated hydrogen yield, are
shown in Fig. 2 (lower panel). The experimental ratios, even though evaluated with large errors, are found to
scatter within the intervals given by the calculations for $\gamma$ = 0.5 and 1.5. Linear interpolations between
the predictions, averaged over $0.3 < p_{t}/A < 1.0$ GeV/c, yield $\gamma = 1.01\pm0.21$ for the exponent describing
the density dependence of the potential term. The same analysis has been also performed for a different
parameterization of the momentum dependence of the elastic in-medium nucleon-nucleon cross section~\cite{QLi10},
for the $v_{2}$ ratios of free neutrons with respect to free protons, and for different bins of impact parameter, resulting in a preliminary value of $\gamma = 0.9\pm0.3$. However, due to
the limited statistic, a more accurate extrapolation of the symmetry energy is not possible.\\
In an independent analysis, Cozma used data from the same experiment and investigated the influence of several
parameters on the difference between the elliptic flows of protons and neutrons using the T\"{u}bingen version of
the QMD transport model~\cite{Coz11}. They included the parametrization of the isoscalar EoS, the choice of
various forms of free or in-medium nucleon-nucleon cross sections, and model parameters as, e.g., the widths of
the wave packets representing nucleons. The interaction used by Cozma contains an explicit momentum dependence of
the symmetry energy part. Both the difference and the ratio of neutron and proton elliptic flows are found to be
sensitive to the density dependence of the symmetry energy; comparing the predictions to the FOPI/LAND data in
~\cite{Coz11}, the best description is obtained with a density dependence dependence slightly stiffer than the $\gamma = 0.5$ case; it is thus rather close to the UrQMD result $\gamma = 0.9$ which may represent an important
step towards the model invariance with respect to different treatment of the nucleon-nucleon cross sections used
in the two studies and to the explicit momentum dependence of the isovector potentials which is implemented in
the T\"{u}bingen QMD but not in the UrQMD version.
\\However, it is obvious that data with higher statistical precision are highly desirable, since it allows following the
evolutions of isospin signals with impact parameter, transverse momentum and particle type more accurately.
Moreover better data would be important to study isospin effects on the momentum dependence of the in-medium
interactions \cite{Gio11}.

\section{ASY-EOS (S394) experiment at GSI}
In May 2011 the data taking of experiment S394 at GSI has been completed. The symmetric collision systems
$^{197}Au+^{197}Au$, $^{96}Zr+^{96}Zr$ and $^{96}Ru+^{96}Ru$ at 400 MeV/nucleon incident energies have been
measured. With the new experiment, an attempt is being made to considerably improve the previous set of data, by
improving the statistical accuracy of the measured flow parameters for Au+Au reactions, and to extend the flow
measurements to other systems. Indeed the study of isospin effects can be improved using new observable like the
one related to light fragments up to atomic number Z=4, with special emphasis on the light isobar pairs
$^{3}H/^{3}He$ and $^{7}Li/^{7}Be$.\\
A schematic view and a photo of experimental set-up are shown in Fig. 3.
\begin{figure}
 \includegraphics[scale=.25]{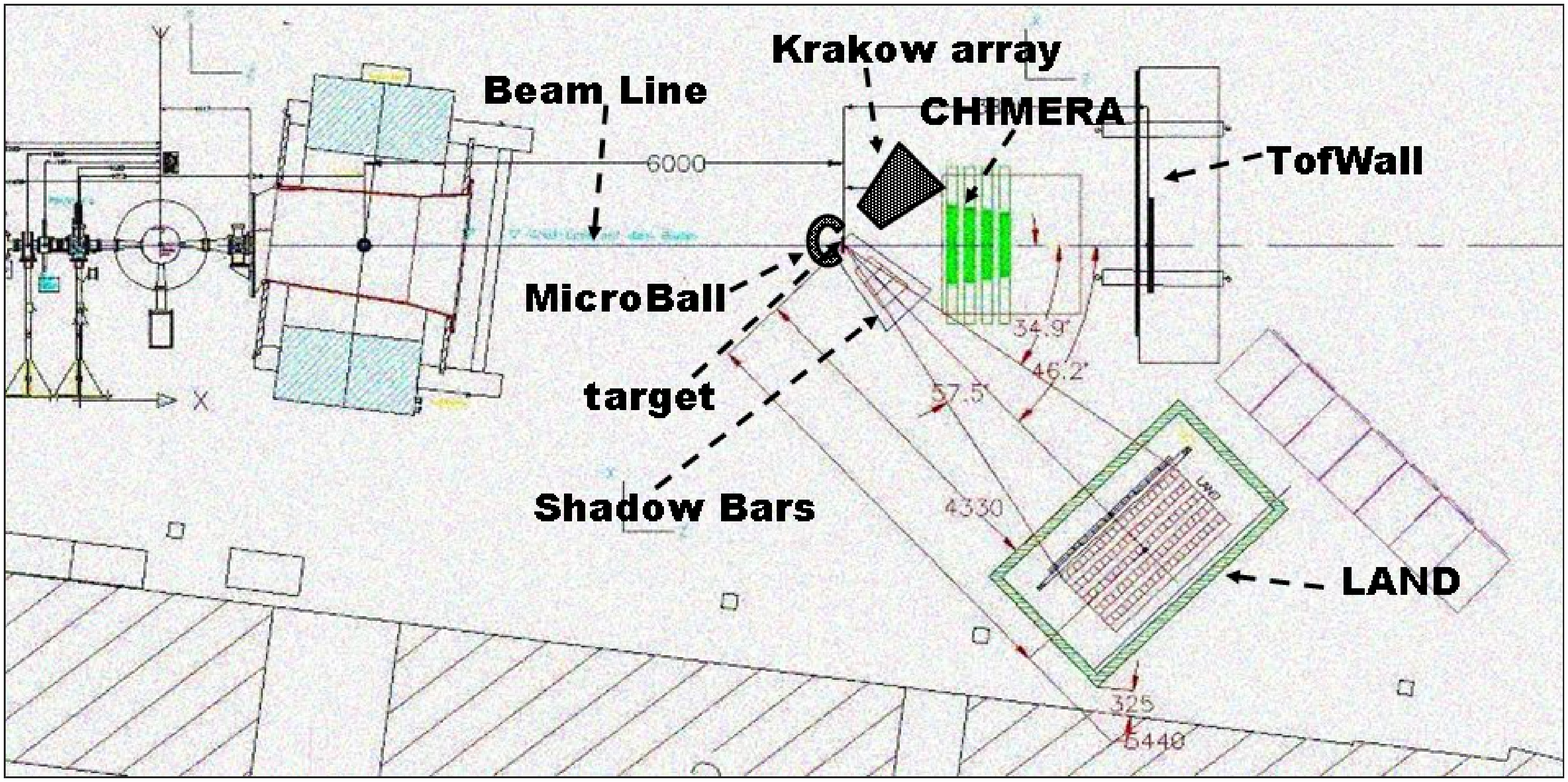}\includegraphics[scale=.3]{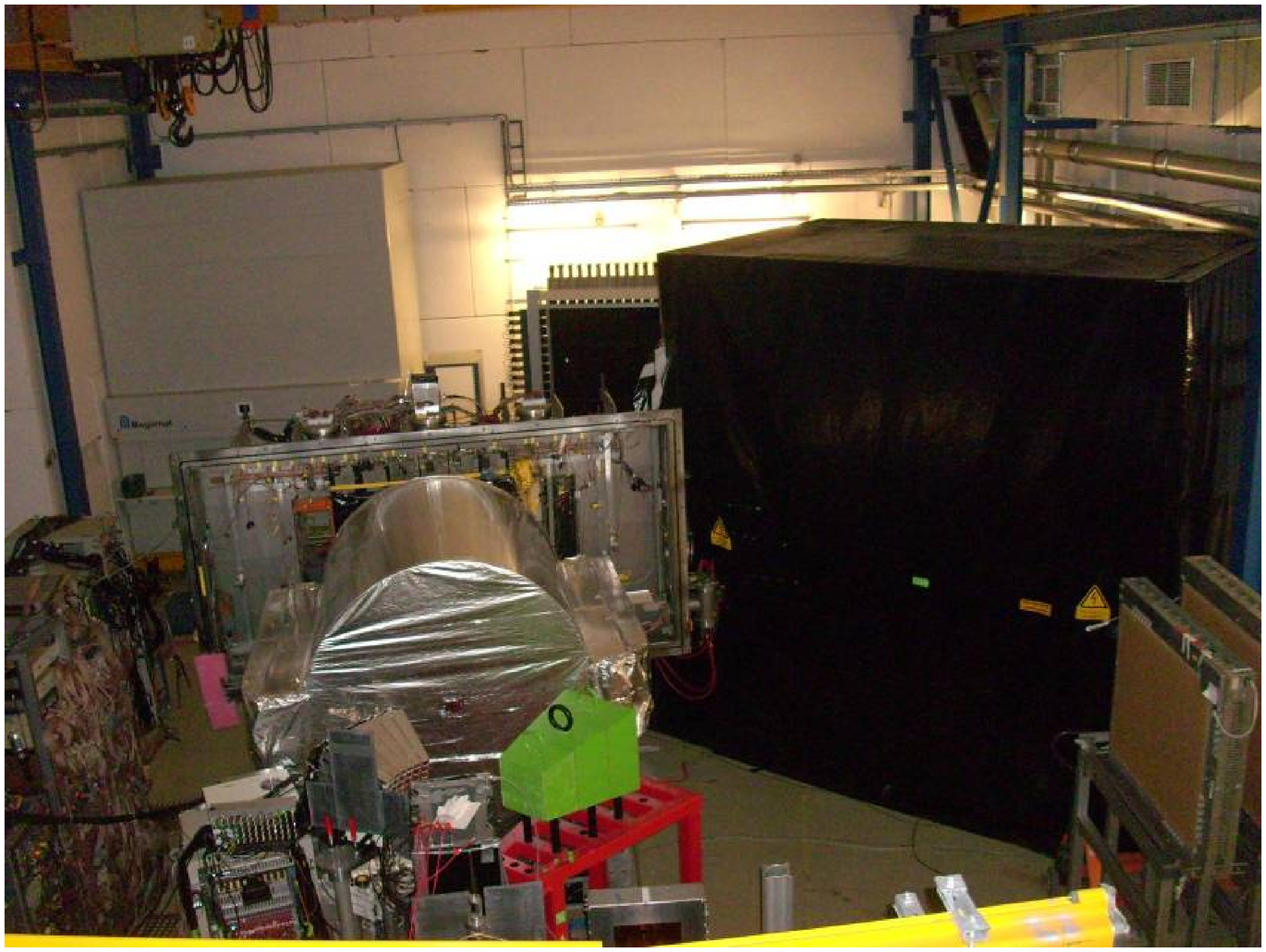}
\caption{Floor plan (left) and view in beam direction (right) of the experimental
setup showing the covered CHIMERA (cylinder), the frame of the ToF-Wall (behind) and
LAND (large black cube to the right).}
\label{fig03} 
\end{figure}
The beam was guided in vacuum to about 2 m upstream from the target. A thin plastic foil read by two
photo-multipliers was used to tag in time the beam arrival and acted as a start detector for the time-of-flight
measurement.
\\The Large Area Neutron Detector (LAND) \cite{LAND}, recently upgraded with new TACQUILA GSI-ASIC electronics,
was positioned at laboratory angles around 45$^{\circ}$ with respect to the beam direction, at a distance of about
5 m from the target. A veto-wall of plastic scintillators in front of LAND allows discrimination of neutrons and
charged particles. In such a way it is possible to measure the direct and elliptic collective flows of neutrons
and hydrogens at mid-rapidity with high precision in the same angular acceptance.\\
In addition, Krakow Triple Telescope Array, KraTTA \cite{Luk11}, has been built to measure the energy, emission
angles and isotopic composition of light charged reaction products. It performed very well during
the experiment. The 35 modules of KraTTA, arranged in a 7x5 array, were placed opposite to LAND at the distance of
40 cm from the target, and covered 160 $msr$ of the solid angle, at polar angles between 20$^{\circ}$ and
64$^{\circ}$. The modules of KraTTA, shown in panel a) of Fig. 3 consisted of two, optically decoupled, CsI(Tl)
crystals (thickness of 2.5 and 12.5 $cm$) and three large area, 500 $\mu$m thick, PIN photo-diodes. The first
photo-diode served as a Si $\Delta$E detector and supplied the ionization signal alone. The second one worked in a
“Single Chip Telescope” configuration and provided both the ionization signal and the light output from the thin
crystal. The third photo-diode read out the light from the thick crystal. The signals from the photo-diodes were
integrated by custom-made low-noise charge preamplifiers and digitized with 100 MHz, CAEN V1724 digitizers.
Very good isotopic resolution has been obtained in the whole dynamic range (see \cite{Luk11}). As an example panel
b) of Fig. 4 shows a $\Delta$E-E scatter plot from the KraTTA detector; an isotopic resolution up to Z=6 is clearly
visible.\\The determination of the impact parameter and the orientation of the reaction plane required the use of several
devices:\\
i) the ALADIN Time-of-Flight wall \cite{TW} was used to detect forward emitted charged particles at polar
angles smaller than 7$^{\circ}$; two  walls (front and rear) of 2.5*100 cm$^{2}$ plastic scintillators, read by
two photo-multipliers at both ends, gave information on emission angle, atomic number and velocity of ions. A
Time Of Flight vs $\Delta$E scatter plot is shown in panel d) of Fig. 3, for a single scintillator; the different
elements can be easily separated. The obtained time of flight resolution, with respect to the start detector,
varied with Z, smoothly decreasing from $\sim$ 250 ps (standard deviation) for lithium fragments to $\sim$ 100 ps for fragments with Z$\geq$ 10 \cite{Ogu11}.\\
ii) 50 thin ($\sim$ 1 cm thick) CsI(Tl) elements, read out by photo-diodes, arranged in 4 rings of the
Washington-University $\mu$-ball array \cite{BALL}, covering polar angles between 60$^{\circ}$ and 147$^{\circ}$,
surrounded the target with the aim of measuring the distribution of backward emitted particles and to discriminate
against background reactions on non-target material;\\
iii) 352 CsI(Tl) scintillators, 12 cm thick, of the CHIMERA multidetector \cite{Pag04}, arranged in 8 rings in
$2\pi$ azimuthal coverage  around the beam axis, covered polar angles between 7$^{\circ}$ and 20$^{\circ}$,
measuring the emission of light charged particles. In addition, thin (300 $\mu$m) Silicon detectors were placed in front of 32
(4 by ring) CsI detectors in the usual $\Delta$E-E configuration.
Identification of particles in CHIMERA CsI(Tl) has been performed using Pulse Shape Analysis based on standard
fast-slow techniques; an example is shown in panel c) of Fig. 4. We have obtained isotopic identification for
p,d,t and $^{3,4}He$ ions stopped in the CsI detectors.
Particles punching through the CsI(Tl) can be distinguished from stopped particles and identified, however only in atomic number Z; the difference in ionization densities $dE/dx$ between stopped and punching through ions results in a sufficiently different fast/slow ratio.
At the lowest fast and slow values an intense ridge due to gamma, fast electrons, neutrons interaction in the
detector medium and background reactions on non-target material is found.\begin{figure}[t]
\includegraphics[scale=.72]{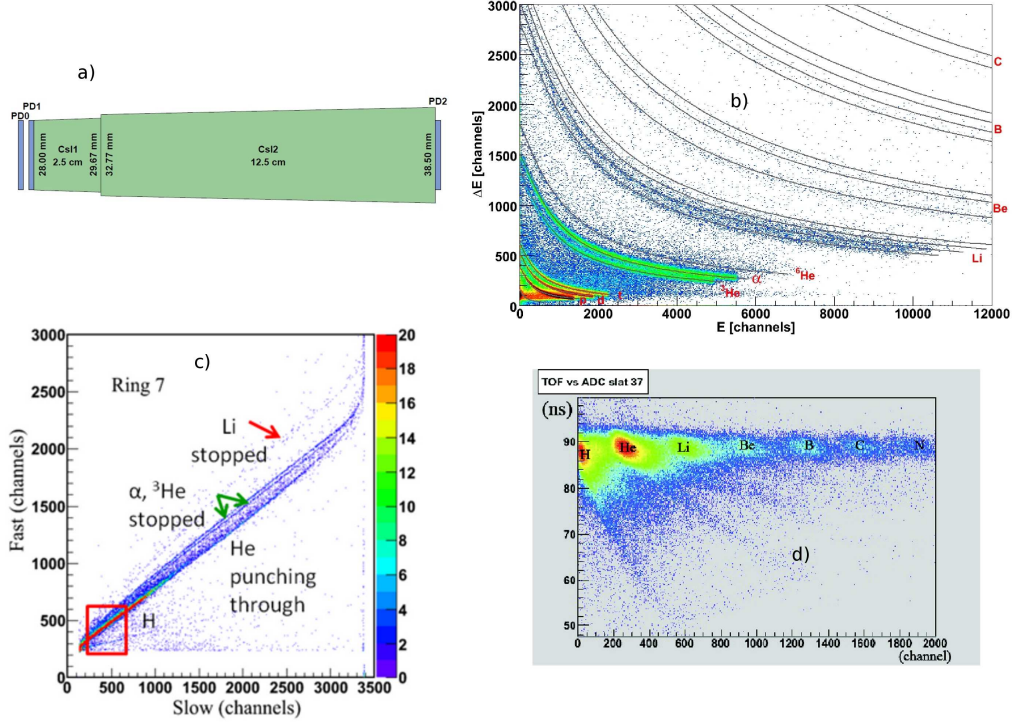}
\caption{Schematic view of a KraTTA module (a) and scatter plots from Au+Au reactions at 400 MeV/nucleon: $\Delta$E-E (CsI vs. CsI) from KraTTA (b); fast-vs.-slow components from a CHIMERA CsI(Tl) scintillator
placed at a polar angle $\theta_{lab}\sim17^{\circ}$ (c); time-of-flight vs. $\Delta$E from a ToF-Wall
detector (d).
}
\label{fig04} 
\end{figure}
For particles heavier than helium, the slow component is partially saturated, since the gate width for the CsI
slow component has been chosen in order to compromise between a good separation of hydrogen isotopes and
identification of Li/Be ions within the codifier's maximum energy range. The identification in CsI has been
cross-checked with the one obtained in the 32 Si-CsI telescopes via $\Delta$E-E technique. In addition, a digital
acquisition sampling technique (14 bit, 50 MHz sampling frequency) was used, in parallel to the standard analog one, in
about 10 $\%$ of the detectors. Cross checking of identification between the standard analog and the digital DAQ has
been of fundamental importance; more results are given in ~\cite{Gua11}.
Energy calibration of the fast component has been performed via the evaluation of the punching through points. For
particles punching through the detectors, the total kinetic energy has been evaluated from the measured $\Delta$E
using energy-loss tables.\\
With beam intensities of about $10^{5}$ pps and targets of 1-2$\%$ interaction probability, about $5*10^{6}$
events for each system were collected. Special runs were performed with and without target, in order to measure
the background from interaction of projectile ions with air, and with iron shadow bars covering the angular
acceptance of LAND in order to measure neutron background. Data acquisition was performed using the MBS data
acquisition system available at GSI \cite{MBS}; the CHIMERA data acquisition was integrated into the MBS system
using the time-stamping technique for data synchronization.
\begin{figure}[t]
\includegraphics[scale=.50]{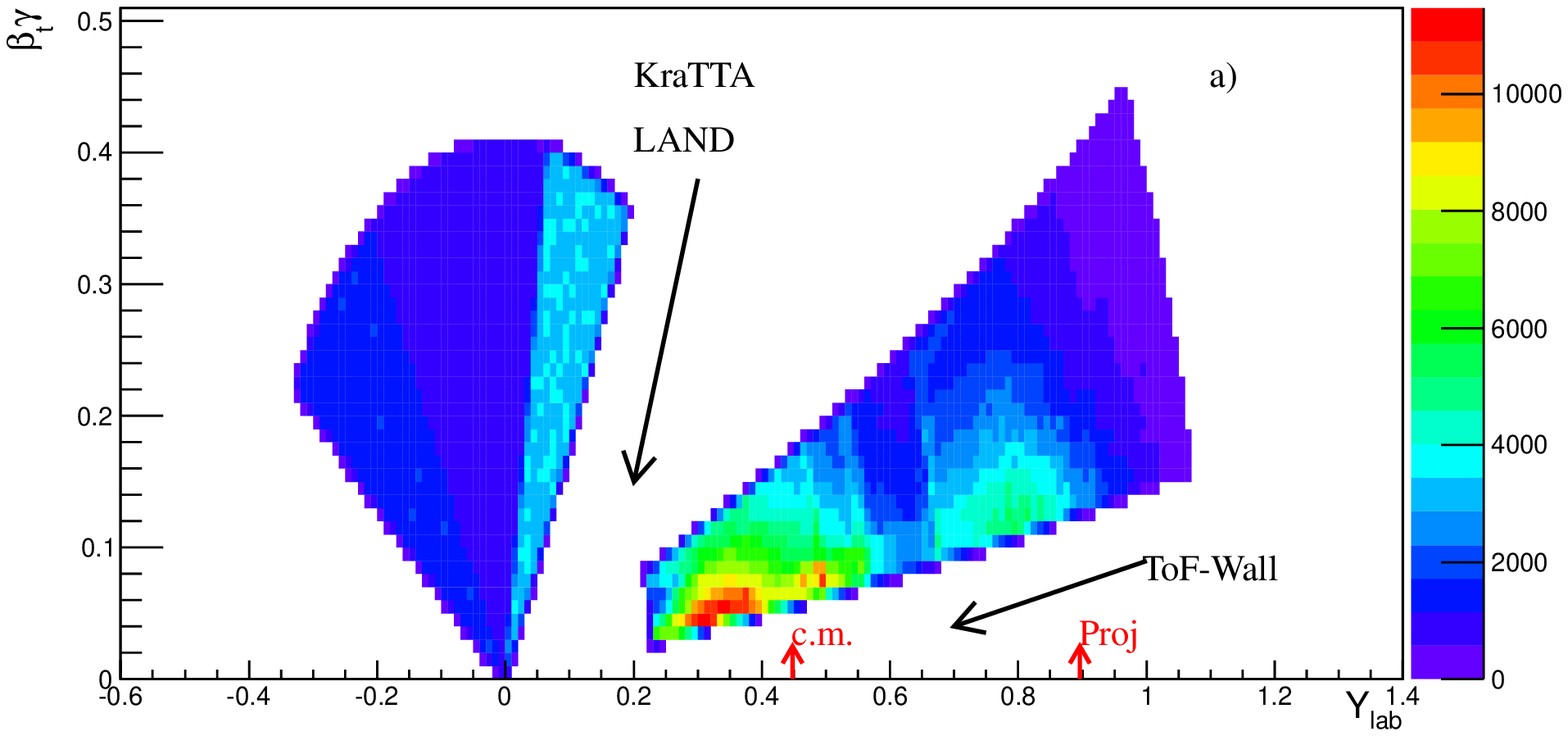}
\includegraphics[scale=.25]{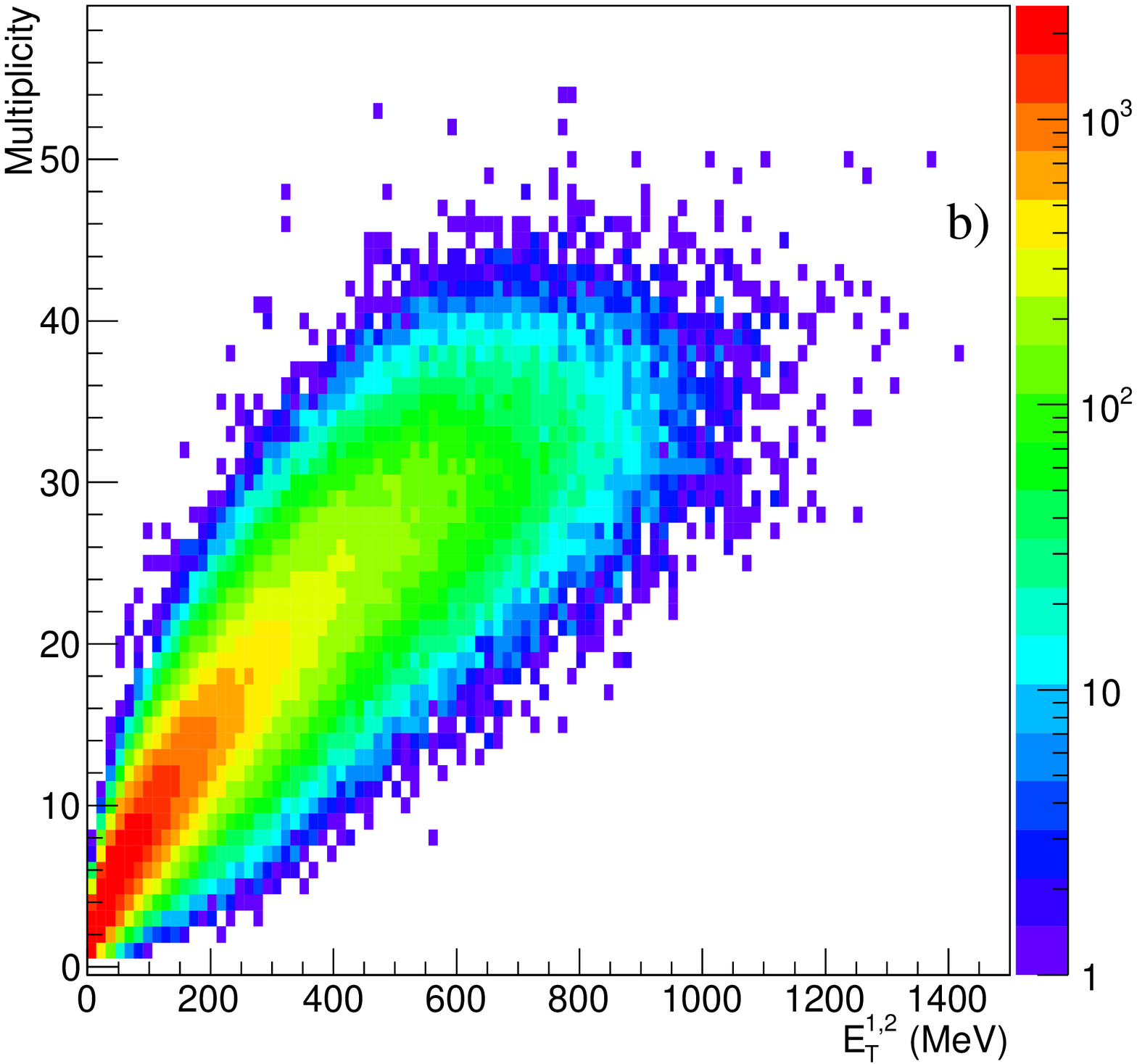}
\caption{panel a): transverse velocity versus rapidity in lab reference system for particles detected by CHIMERA
and $\mu$-ball array for Au+Au data at 400 MeV/nucleon; panel b): correlation between  total transverse energy of
light charged particles and multiplicity in CHIMERA.
}
\label{fig05} 
\end{figure}
The analysis of the collected data has been started with calibrations of the individual detector systems and with
overall quality checks, and is currently in progress. At the present stage we will report here only on some
preliminary results.\\
As a global result we show in panel a) of Fig. 5 the transverse velocity vs rapidity in the lab reference system
for the Au+Au system, for particles detected by CHIMERA and $\mu$-ball detector. In the case of $\mu$-ball, since
energy calibration is not available, we simulated a uniform kinetic energy distribution between 0 and 75 MeV; in
the figure we can clearly see population of two intense regions around mid-rapidity and projectile rapidity. In
order to reject fast electrons and background, a threshold of E/A$>$25 MeV/nucleon has been imposed in CHIMERA.\\
It is interesting to look at the the evolution of kinematic distributions with increasing collision violence; as  a
first test, we have used the total transverse energy of light charged particles, $E_{T}^{12}=\sum_{i}
E_{Kin}^{i}*sin^2(\theta_{i}) (Z_{i}\leq 2)$ as an estimator of collision violence. The correlation between
$E_{T}^{12}$ and total charged particle multiplicity in CHIMERA is shown in panel b) of Fig. 5.
The $1^{st}$ row of Fig. 6 shows the evolution of the transverse velocity vs. rapidity scatter plots for particles
detected by CHIMERA for different selections of $E_{T}^{12}$; the collision violence increases from left to right.
An important parameter is the resolution achieved in determining the azimuthal orientation of the reaction plane,
that largely determines the uncertainty associated with the determined flow parameters \cite{And08}.
\begin{figure}[t]
\includegraphics[scale=.8]{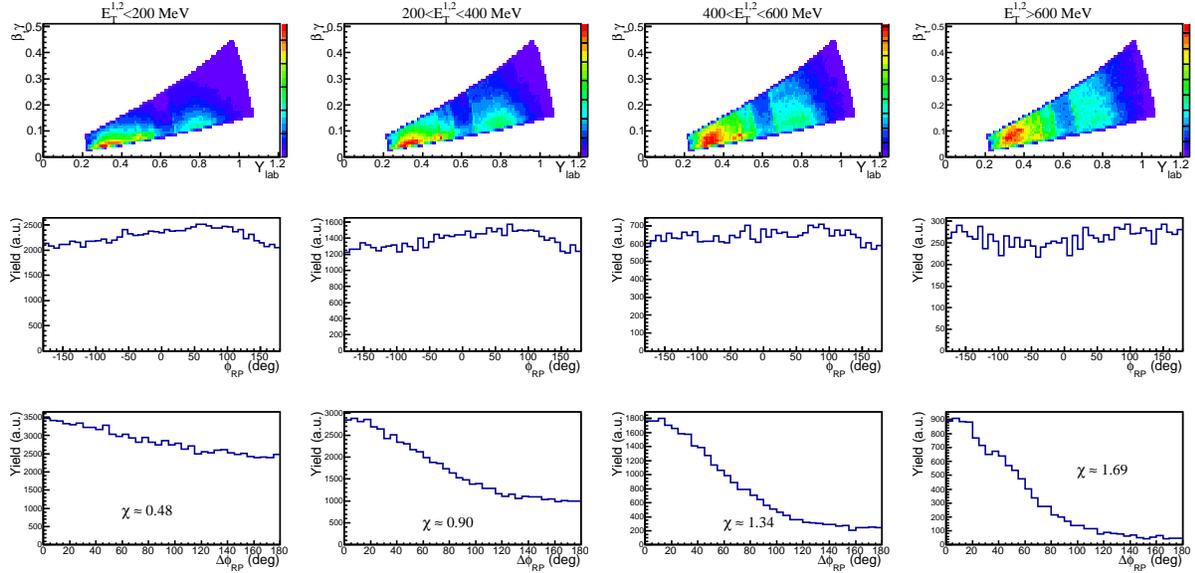}
\caption{CHIMERA data for Au+Au system; $1^{st}$ row: evolution of transverse velocity versus rapidity in lab
reference system for different selection of $E_{T}^{12}$; $2^{nd}$ row: orientation of reaction plane obtained
using Q-vector method for different selection of $E_{T}^{12}$; $3^{rd}$ row: difference of orientations of
reaction plane as obtained using sub-events mixing technique. Value of the
reaction plane dispersion parameter $\chi$ (see \cite{Ollxx}) are reported. }
\label{fig06} 
\end{figure}
As a first test we have estimated the reaction plane orientation for events with total charged particle
multiplicity M$\geq$4 in CHIMERA, using the Q-vector method of Ref. \cite{Dan}. In order to reject the
mid-rapidity region, a cut on laboratory rapidity $Y_{lab}>$0.548 (corresponding to $Y>$0.1 in c.m. system) was used.
\begin{figure}[]
\includegraphics[scale=.60]{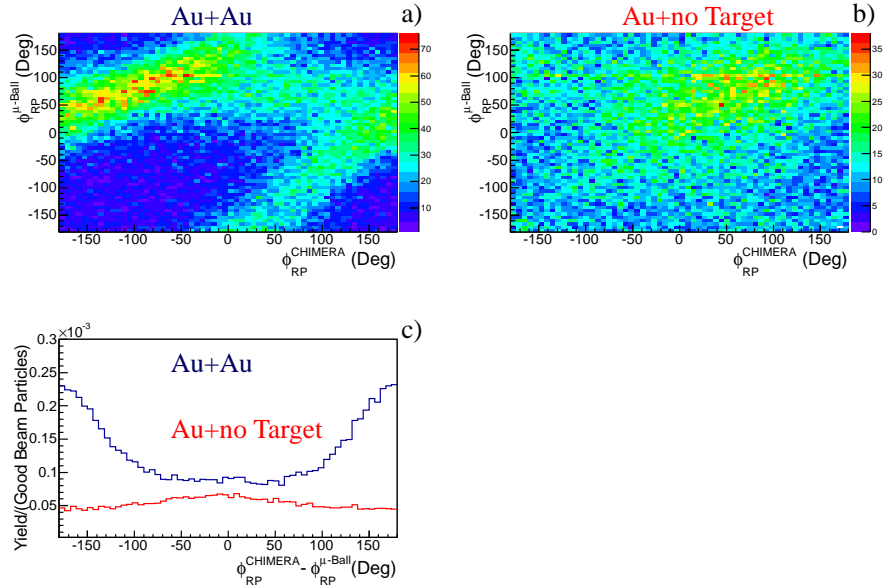}
\caption{Correlation between reaction plane orientation as given by CHIMERA (x axis) and $\mu$-ball (y axis) for
for Au+Au data (panel a) and for Au+no target data (panel b); difference of the two reaction plane orientations
(CHIMERA minus $\mu$-ball) for Au+Au data  and for Au+no target data (panel c), normalized to the integrated beam
intensity.
}
\label{fig07} 
\end{figure}
The obtained reaction plane distribution for a CHIMERA data sample from the Au+Au data set, for different
selection of $E_{T}^{12}$,  is shown in $2^{nd}$ row of Fig. 6; the flatness indicates that the particle angular
distributions have not been biased by the event triggering in the experiment. We also tested the resolution
achieved in reconstructing the reaction plane using the sub-event mixing technique of Ref. \cite{Ollxx}. The
distribution of the difference between the two reaction plane orientations extracted by the sub-events is reported
in the $3^{rd}$ row of Fig. 6.
Using the method of \cite{Ollxx} we obtain a reaction plane dispersion parameter $\chi\sim $0.90 for
$200<E_{T}^{12}<400 MeV$, resulting in an attenuation of the elliptic flow measurement of $\sim$ 0.34, and a
$\chi\sim $1.7 for $E_{T12}>600 MeV$, corresponding to an attenuation of $\sim$ 0.66. this has been estimated from Fig. 4 of \cite{Ollxx} which shows correction factors for the Fourier parameters of the azimuthal distribution v$_{n}$ of order n as a
function of $\chi$. In the case of low dissipative collisions ($E_{T}^{12}<200 MeV$), most of the particles from projectile fragmentation are hitting the ToF-Wall detector. The very low multiplicities ($<\sim 4$) of particles detected by CHIMERA do not permit an accurate determination of the reaction plane orientation.
However, the analysis performed so far shows that the reaction plane orientation achieved with the CHIMERA
detector modules is better than the one estimated in simulations. It is to be expected that
these values will improve considerably as soon as the information collected with the Time-of-Flight wall can be
included in the analysis.
In Fig. 7 we show the correlation between reaction plane orientations as determined by CHIMERA and $\mu$-ball subset
of data for Au+Au reactions (panel a) and Au+no target data (panel b); a nice anti-correlation is found for
on-target reaction, since the two devices detect particles in the opposite hemispheres (projectile and target);
the data collected without target shows a spurious background. The difference between the two reaction plane
orientations, normalized to the integrated beam intensity, is presented in panel c), showing how the reaction
on non target materials can be discriminated by on-target reaction; this proves the usefulness of $\mu$-ball data
in rejecting the background reactions.
\section{Acknowledgements}
Work supported by EU under contract No. FP7-25431 (Hadron-Physics2). One of us (S.K.) acknowledges support by the Foundation for Polish Science-MPD program, co-financed by the European Union within the European Regional Development Fund.

\end{document}